\documentclass[twocolumn,preprintnumbers,amsmath,amssymb,aps,prb]{revtex4}
\usepackage{graphicx}
\begin{document}

\title{Disordering, Clustering, and Laning Transitions in Particle Systems with Dispersion in the Magnus Term}
 
\author{C. J. O. Reichhardt and C. Reichhardt}
\affiliation{Theoretical Division and Center for Nonlinear Studies,
Los Alamos National Laboratory, Los Alamos, New Mexico 87545, USA}

\date{\today}

\begin{abstract}
We numerically examine a two-dimensional system of repulsively interacting particles with dynamics that are governed by both
a damping term and a Magnus term.  
The magnitude of the Magnus term has one value for half of the particles and a different value for the other half of the particles.
In the absence of a driving force,
the particles form a triangular lattice, while
when a driving force is applied, we find that there is a
critical drive above which a
Magnus-induced disordering transition 
can occur even if the
difference in the Magnus term
between the two particle species is as small as one percent.
The transition arises due to the  
different Hall angles of the two species, which causes their motion
to decouple at the critical drive.
At higher
drives, 
the disordered state can
undergo both species and density phase separation
into a
density modulated stripe that is oriented perpendicular to the driving direction.
We
observe several additional phases that occur as a function of
drive and 
Magnus force
disparity, including
a variety of density modulated diagonal laned phases. 
In general we find a much richer variety of states compared to
systems of oppositely driven overdamped Yukawa particles.
We
discuss the implications of our work for 
skyrmion systems,
where we predict that even for small skyrmion dispersities,
a drive-induced disordering transition
can occur along with clustering phases and pattern forming states.       
\end{abstract}

\maketitle

\vskip 2pc

\section{Introduction}

There is a wide variety of systems that can be
effectively modeled as an assembly of interacting particles 
that
undergoes structural transitions under some form of  external driving.
In the presence of a random or periodic substrate,
the particles can exhibit a depinning transition \cite{1,2}
such as that found for
superconducting vortices \cite{3,4}, colloidal particles \cite{5,6,7}, 
or
sliding friction \cite{8}.
Disordering or ordering transitions can occur 
in the absence of quenched disorder
as a function of dc or ac shearing \cite{9,10,11,12,13}.   
In many cases, the particles
have a uniform size and particle-particle interaction force,
but when
the particle sizes or interactions become polydisperse,
order-disorder transitions can appear even in the absence
of driving or shearing
\cite{14,15,16}. 
Disordering transitions and other dynamical phases
can also arise in systems with monodisperse
particle-particle interactions
if some of the particles have different dynamics than others.
A well-studied example
is oppositely driven repulsive particles,
which can form static ordered states 
or undergo fluctuating disordered flow that is followed at higher drives by
a transition
to a laned state consisting of
multiple partially phase separated
oppositely moving stripes \cite{17,18,19,20,21,22,23,24,25,26}.
Similar ordering and laning transitions appear when
the particles move at different velocities in the same direction \cite{27}.
Experimentally, laning transitions
have been realized for colloidal
particles \cite{28,29} and dusty plasmas \cite{30,31}.

Here we study whether a disordering transition or lane formation 
can occur for an assembly of bidisperse particles that are
all driven in the same direction when
each particle species has a different
non-dissipative Magnus term.
We
consider repulsively interacting particles
that form a triangular lattice in the absence of a driving force or substrate.
A dissipative force of magnitude $\alpha_{d}^i$
aligns the velocity of particle $i$ with the net direction of the
external forces acting on that particle,
while a Magnus term of magnitude
$\alpha_{m}^i$
aligns the particle velocity perpendicular to the 
external forces.
When we introduce an applied driving force
of magnitude $F_{D}$, we find that
if the Magnus term $\alpha_{m}^i=0$
and the dissipative term $\alpha_d^i=\alpha_d$
for all $i$,
the system forms a triangular lattice that moves parallel to the driving direction.
If the Magnus term is nonzero but equal for all particles, $\alpha_m^i \equiv \alpha_m$,
a triangular lattice still forms, but it moves
at a Hall angle $\theta_{Sk}=\theta_{Sk}^{\rm int}$ with respect to the driving direction, where 
the intrinsic Hall angle $\theta_{Sk}^{\rm int} = \arctan(\alpha_{m}/\alpha_{d})$.
If the Magnus term is bidisperse, with a value of $\alpha_m^a$ for half of the
particles and $\alpha_m^b>\alpha_m^a$ for the other half, we find that when
$\alpha_m^a=0$ and $\alpha_m^b \neq 0$,
a triangular lattice appears that
moves elastically at an angle
$\theta_{Sk}^{b,\rm int}/2$
for small drives.
Above a critical drive $F_c$,
dislocation pairs proliferate in the lattice and
a dynamical disordering transition occurs
when the two species
move at different velocities in the
directions parallel and perpendicular to the drive
due to the
drive dependence of
$\theta^{b}_{Sk}$.
When $\alpha_m^a \neq 0$ and $\alpha_m^a<\alpha_m^b$, this behavior persists
since the drive dependence of $\theta^a_{Sk}$ differs from that of
$\theta^b_{Sk}$; however, 
$F_{c}$ increases as the difference $\alpha_m^b-\alpha_m^a$ decreases.
At high drives,
the disordered state transitions to a cluster or 
stripe state in which
the particles phase separate into a single stripe oriented perpendicular to the drive
with one species on each side of the stripe.
The stripe becomes denser with increasing $F_{D}$
since
$\theta_{Sk}^b-\theta_{Sk}^a$ increases and
species $a$ piles up behind species $b$.
Due to the increasing compression of the stripe,
eventually an instability occurs
in which the system can lower its
particle-particle interaction energy
by forming a more uniform state
that we call a diagonal laned phase.
In some cases, the diagonal laned state
exhibits strong density modulations
as well as additional transitions to
a larger number of thinner lanes.  All of these
transitions
are associated with changes and jumps in the
average velocity both parallel and perpendicular to the drive, as
well as
changes in the amount of six-fold ordering in the system.   

Our results have implications for driven magnetic skyrmions,
which are particlelike magnetic textures that interact repulsively
with each other and form a triangular lattice \cite{32,33,34}.
Skyrmions can be set into motion by the application of a current
\cite{34,35,36,37,38,39,40}, and
due to the Magnus term
they move at an angle with respect to the driving force 
known as the skyrmion Hall angle \cite{34,41,42,43,44,45}.
Within a given sample, there can be dispersion in the
size of skyrmions, and different species of skyrmions with different dynamics
may be able to coexist with each other
\cite{38,45,L1,L2,N,N2,P}, so it is
important to understand how polydispersity affects
the collective motion of skyrmions.
   
Our results
suggest that if there
is
the slightest dispersion in the Magnus term,
a drive-induced disordering transition from a crystal to a liquidlike state
can occur even in the absence of quenched disorder.
In systems with strong quenched disorder,
monodisperse skyrmions depin
into a plastically flowing disordered state,
but at higher drives they
dynamically reorder into a moving crystal state \cite{41,46,47}
similar to that found for
superconducting vortices \cite{2,3,4,48}.
Our results indicate that when there is a dispersion in the Magnus
term, such behavior is reversed and the skyrmions {\it disorder}
at higher drives.
In addition, 
clustering or species segregation can occur.
Recent continuum and particle-based simulations
of monodisperse skyrmions
showed that
clustering transitions can occur in samples containing
strong pinning or quenched disorder \cite{47,49}.
Our results
demonstrate that clustering
can also occur in the absence of pinning
when there is
any dispersity in the skyrmions
that produces differences in the 
Magnus term.
These results may also be relevant for 
soft matter systems in which Magnus forces are important,
such as magnetic particles in solutions \cite{50,51,52} 
or spinning colloidal particles \cite{53,54},
where different size particles could experience
different effective Magnus forces.

The paper is organized as follows.  In Sec.~II we describe our simulation details.
Section III introduces the Magnus induced disordering transition.  In Sec.~IV we
focus on cluster and stripe formation, and show that a nonequilibrium conformal crystal
structure can spontaneously emerge in the system.
Section V describes the effect of varying the ratio of the damping term for the
two species.  In Sec.~VI we vary the ratio of the Magnus term of the two species for
fixed and equal damping terms.   We show the results of introducing Magnus terms
of opposite sign in Sec.~VII.  In Sec.~VIII we briefly describe the effect of changing
other variables and provide a general discussion.  A summary
of the work appears in Sec.~IX.

\section{Simulation}
We consider a two-dimensional system of size $L \times L$
with periodic boundary conditions in the $x$ and $y$-directions
containing $N_a$ particles of species $a$ and $N_b$ particles of species $b$
for a total of $N=N_a+N_b$ particles.
The particle density is $n=N/L^2$, where $L=36$.
Unless otherwise noted, we take $N_a=N_b=N/2$.
The equations of motion for particle $i$ of species $\gamma=a$ or $b$ is 
\begin{equation} 
\alpha^{\gamma}_d {\bf v}_{i} + \alpha^{\gamma}_m {\hat z} \times {\bf v}_{i} =
{\bf F}^{ss}_{i} + {\bf F}^{D}
\end{equation}
where the particle velocity is
${\bf v}_{i} = {d {\bf r}_{i}}/{dt}$.
All particle-particle interactions have the same 
pairwise
form of a modified Bessel function, ${\bf F}_i^{ss}=\sum_{i}^{N}K_{1}(r_{ij}){\bf {\hat r}}_{ij}$
that falls
off exponentially for large $r$.
Here $r_{ij}=|{\bf r}_i-{\bf r}_j|$ is the distance between particles $i$ and $j$,
and ${\bf {\hat r}}_{ij}=({\bf r}_i-{\bf r}_j)/r_{ij}$.
This interaction potential has been used previously
for particle-based models of skyrmions,
and in the absence of pinning it causes
the particles
to form a hexagonal lattice \cite{37,41,46,49,55}.      
The driving force ${\bf F}^{D}=F_D{\bf {\hat x}}$ is the same
for all particles.
We increase the drive in increments of
$\delta F^D = 0.002$
and wait $10^4$ simulation time steps between increments to ensure we 
are in a steady state.     
The 
damping term
$\alpha^{\gamma}_d$
aligns the particle velocity
in the direction of the net applied forces, while
the
Magnus term
$\alpha^{\gamma}_{m}$
generates velocity components that 
are perpendicular to the net applied forces. 
As in previous work,
unless otherwise noted we
normalize the two coefficients such that
$(\alpha_{d}^{\gamma})^2 + (\alpha_{m}^{\gamma})^2 = 1.0$ \cite{41,42,55};
however, we also consider
systems with fixed $\alpha_{d}^{\gamma}$ and varied $\alpha_{m}^{\gamma}$ that
are not subject to this constraint.
The intrinsic Hall angle for species $\gamma$ is
$\theta^{\gamma,{\rm int}}_{Sk} = \arctan(\alpha_{m}^{\gamma}/\alpha_{d}^{\gamma})$.
It is known from previous work on skyrmion systems with disorder
that  $\theta_{Sk}^{\gamma}$ depends on
the velocity of the particles and can be written
as $\theta_{Sk}^{\gamma} = \tan^{-1}(\langle V_{\perp}^{\gamma}\rangle/\langle V_{||}^{\gamma}\rangle)$
where
$\langle V_{\perp}^{\gamma}\rangle=N_{\gamma}^{-1}\sum_i^{N_{\gamma}}{\bf v}_i \cdot {\bf {\hat y}}$ and
$\langle V_{||}^{\gamma}\rangle=N_{\gamma}^{-1}\sum_i^{N_{\gamma}}{\bf v}_i \cdot {\bf {\hat x}}$
\cite{41,42,46,56}.
In a system
with $\alpha^{a}_{m}=\alpha^b_m= 0$, $\theta_{Sk}^a = \theta_{Sk}^b = 0$.
We initialize the particles in a triangular lattice and assign $N_a$ randomly selected
particles to be species $a$, with the remaining particles set to species $b$.
We measure
$\langle V_{\perp}^a\rangle$,
$\langle V_{||}^a\rangle$,
$\langle V_{\perp}^b\rangle$, and
$\langle V_{||}^b\rangle$, along with
$\theta^{a}_{Sk}$ and
$\theta^{b}_{Sk}$.

\section{Magnus Induced Disordering Transition}

\begin{figure}
\includegraphics[width=3.5in]{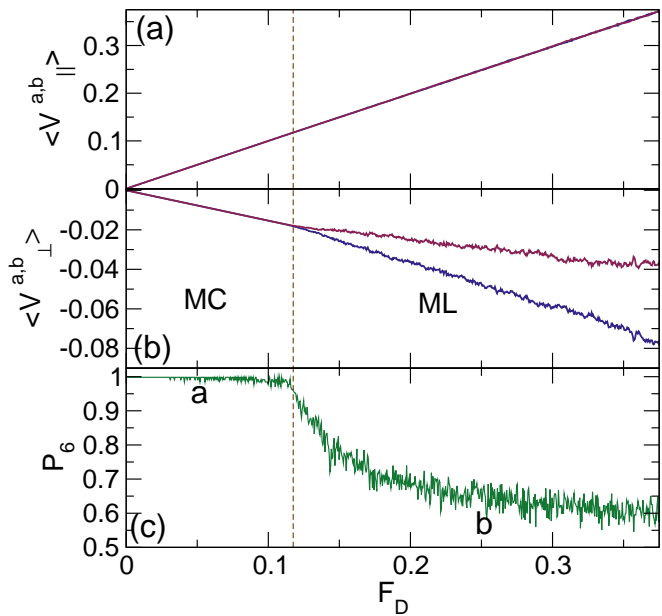}
\caption{$\langle V^a_{||}\rangle$ (red) and $\langle V^b_{||}\rangle$ (blue)
  vs $F_{D}$ in a sample with $\alpha^a_m=0$, $\alpha^a_d=1$,
  and $\alpha^b_m=0.3$.
  (b) The corresponding
  $\langle V^a_{\perp}\rangle$ (red) and $\langle V^b_{\perp}\rangle$ (blue)
  vs $F_{D}$.
  For $F_{D} < 0.1175$, the system forms a moving triangular crystal (MC) and the
  velocities are locked in both directions,
  while for
  $F_{D} \geq 0.1175$,
  the transverse velocity curves
  split with $\langle V^{b}_{\perp}\rangle$ increasing more rapidly
  with $F_D$ than $\langle V^a_{\perp}\rangle$,
indicating that the 
two species are now moving at different velocities. 
(c) The fraction $P_6$ of sixfold-coordinated particles
vs $F_{D}$.
For $F_{D} < 0.1175$, $P_{6} \approx 1$
as expected for a triangular lattice,
while $P_6$ drops for $F_{D} \geq 0.1175$,
indicating
a disordering of the system.
The letters a and b
indicate the values of $F_{D}$ at which
the images
in Fig.~\ref{fig:2} were obtained.
The vertical dashed line marks the transition from the
moving crystal (MC) to the moving liquid (ML) state.}
\label{fig:1}
\end{figure}

\begin{figure}
\includegraphics[width=3.5in]{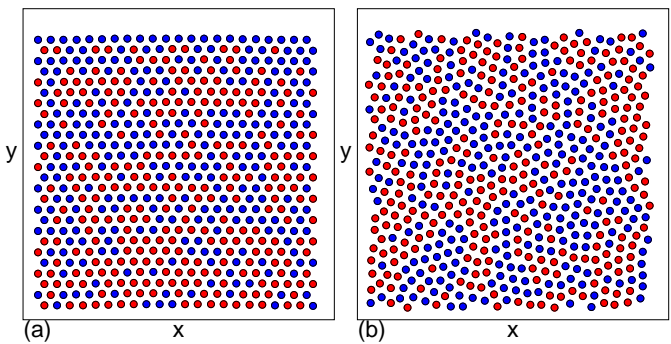}
\caption{Images of particle positions for species $a$ (red) and $b$ (blue) for
  the system in Fig.~\ref{fig:1} with $\alpha_m^a=0$, $\alpha_d^a=1.0$,
  and $\alpha_m^b=0.3$
  at the drives marked a and b in Fig.~\ref{fig:1}(c).
  (a) At $F_{D} = 0.05$,
  the system forms a triangular solid that is moving in the positive $x$ and 
  negative $y$ directions.
  (b) At $F_{D} = 0.25$, the system is in a moving liquid phase.} 
\label{fig:2}
\end{figure}

\begin{figure}
\includegraphics[width=3.5in]{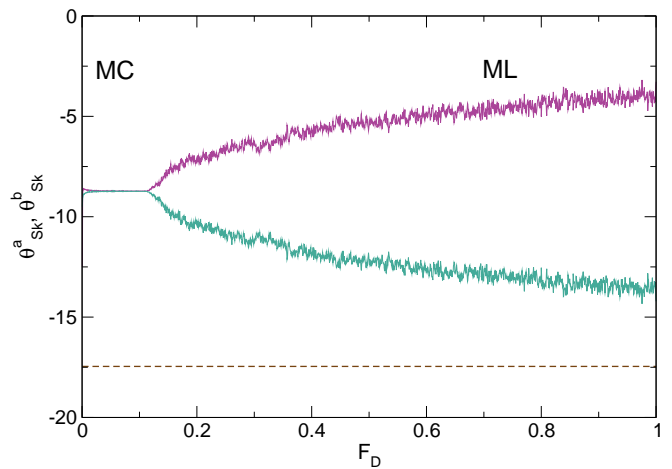}
\caption{ The Hall angles $\theta^a_{Sk}$ (pink) and $\theta^b_{Sk}$ (green)
  vs $F_{D}$ for the system in Fig.~\ref{fig:1} with $\alpha_m^a=0$,
  $\alpha_d^a=0$, and $\alpha_m^b=0.3$.
In the moving crystal (MC) phase, the species are locked together and all the particles 
have $\theta^a_{Sk}=\theta^b_{Sk} = -8.73^\circ$,
while in the disordered moving liquid (ML) state,
the magnitude of $\theta^{a}_{Sk}$ decreases toward $0^\circ$ while
$\theta^{b}_{Sk}$ gradually
approaches $\theta^{b,{\rm int}}_{Sk}=-17.45^{\circ}$,
marked by a dashed line.}    
\label{fig:3}
\end{figure}

We first consider a
system
containing 
$N = 572$ particles at a density of
$n= 0.4413$  which forms a triangular solid
when $F_D=0$.
We fix species $a$ in the overdamped limit
with $\alpha^{a}_{m} = 0$ and 
$\alpha^{a}_{d} = 1.0$,
while species $b$ has $\alpha^{b}_{m} = 0.3$ and
$(\alpha^b_m)^2+(\alpha^b_d)^2=1.0$.
In Fig.~\ref{fig:1}(a) we plot $\langle V^a_{||}\rangle$ and
$\langle V^b_{||}\rangle$
versus $F_{D}$,
and in Fig.~\ref{fig:1}(b) we show the corresponding
$\langle V^a_{\perp}\rangle$ and $\langle V^b_{\perp}\rangle$ versus $F_D$ curves.
In Fig.~\ref{fig:1}(c) we plot $P_6$, the overall fraction of sixfold coordinated
particles, versus $F_D$.  Here $P_6=N^{-1}\sum_i^N \delta(z_i-6)$, where $z_i$ is
the coordination number of particle $i$ obtained from a Voronoi tessellation.
When $F_{D} < 0.1175$, we have
$\langle V^{a}_{\perp}\rangle/\langle V^b_{\perp}\rangle = \langle V^a_{||}\rangle/\langle V^b_{||}\rangle = 1.0$ and $P_{6} = 1.0$, indicating that
the system forms a triangular lattice which moves elastically  
in the positive $x$ and negative $y$ directions.
We illustrate the particle positions at $F_D=0.05$ in
Fig.~\ref{fig:2}(a),
where a moving crystal (MC) phase appears. 
At $F_{D} = 0.1175$,
a disordering transition occurs that is associated
with a drop in $P_{6}$ caused by the proliferation of 5-7 defect pairs.
At this same drive, the $\langle V^a_{\perp}\rangle$ and $\langle V^b_{\perp}\rangle$
curves in Fig.~\ref{fig:1}(b) split, and
the $\langle V^{b}_{\perp}\rangle$ curve
increases more rapidly with $F_D$ than the
$\langle V^a_{\perp}\rangle$ curve,
indicating that the disordering transition is triggered by
a partial decoupling of the two species,
which now move at different
transverse velocities.
In Fig.~\ref{fig:2}(b) we show the particle positions at $F_{D} = 0.25$ 
where the system
has disordered but the particle species
remain mixed.
In this moving liquid (ML) phase, the particles undergo continual  
dynamical rearrangements.
Despite the fact that
$\alpha^{a}_{m} = 0$,
$\langle V^a_{\perp}\rangle$ continues to increase with increasing
$F_D$ in the ML phase, indicating that although the two species are no longer
fully coupled, species $b$ is able to drag species $a$ in the transverse direction.
The vertical dashed line in Fig.~\ref{fig:1}
marks the MC-ML transition
and shows that the longitudinal velocities
$\langle V^a_{||}\rangle$ and $\langle V^b_{||}\rangle$
in Fig.~\ref{fig:1}(a) are not affected by the transition.
In Fig.~\ref{fig:3} we plot the drive dependent Hall angles
$\theta^a_{Sk}$ and $\theta^b_{Sk}$ versus $F_{D}$
for the system in Fig.~\ref{fig:1},
where
the intrinsic Hall angles are
$\theta^{a,{\rm int}}_{Sk}=0^{\circ}$ and
$\theta^{b,{\rm int}}_{Sk}=-17.45^{\circ}$.
In the MC state, both species are locked to the
same Hall angle,
$\theta^a_{Sk}=\theta^b_{Sk} \approx \theta^{b,{\rm int}}_{Sk}/2 = -8.73^{\circ}$,
and at the disordering transition,
$\theta^a_{Sk}$ approaches 
$0^{\circ}$ while
$\theta^b_{Sk}$
approaches its intrinsic value of
$\theta^{b,{\rm int}}_{Sk}=-17.45^\circ$.

\begin{figure}
  \includegraphics[width=3.5in]{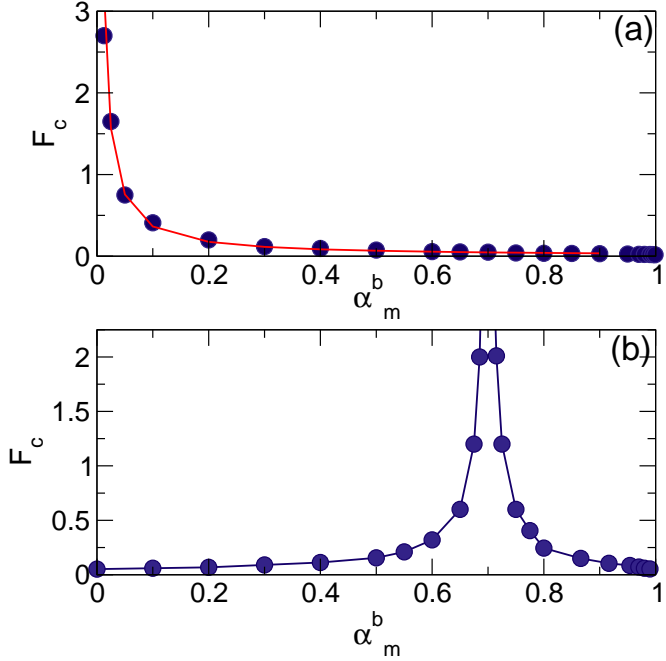}
\caption{(a) $F_{c}$, the drive at which the MC-ML transition occurs,
  vs $\alpha^{b}_{m}$
  for the system in Fig.~\ref{fig:1} with $\alpha^{a}_{m} = 0$ and $\alpha^a_d=1.0$.
  The solid line is a power law fit to 
 $F_c \propto C(\alpha^b_{m})^\beta$ with $C = 0.0323$ and $\beta = -1.05$.
  (b) $F_c$ vs $\alpha^b_m$ in a system with fixed
  $\alpha^{a}_{m} = 0.7$ for
  $\alpha^a_d=\alpha^b_d=1.0$, showing a divergence at 
$\alpha^{b}_{m}/\alpha^{a}_{m} = 1.0$.}  
\label{fig:4}
\end{figure}

In Fig.~\ref{fig:4}(a) we plot the critical force $F_c$, equal to the drive at which
the MC-ML transition occurs, versus
$\alpha^b_{m}$ for the system in Fig.~\ref{fig:1}
with $\alpha_m^a=0$ and $\alpha_d^a=1.0$.
As $\alpha^b_{m}$ increases, the disordering transition shifts to lower drives,
and we find that the critical force can be fit to
$ F_{c}  \propto C (\alpha^{b}_{m})^\beta$ with
$C = 0.0323$ and $\beta = -1.05$. 
These results show that even when the Magnus term is very small,
application of an external drive can
induce a disordering transition.
We note  that under our imposed normalization constraint,
$\alpha_m^b<1.0$.
In skyrmion systems, all the particles have a finite Magnus term,
so in Fig.~\ref{fig:4}(b) we plot $F_{c}$
versus $\alpha^b_{m}$ for a system in which
we vary $\alpha^b_m$ while fixing $\alpha^a_m=0.7$ with $\alpha^a_d = \alpha^b_d=1.0$.
Here $F_{c}$ diverges 
at $\alpha^{b}_{m}/\alpha^a_{m} = 1.0$
according to $F_{c} \propto |\alpha^{a}_{m} - \alpha^b_{m}|^{-1}$.

\section{Cluster and Stripe Formation}

\begin{figure}
\includegraphics[width=3.5in]{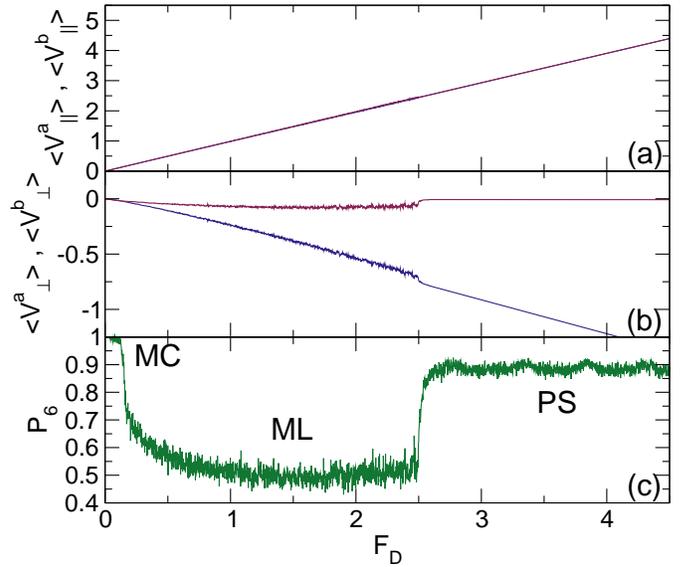}
\caption{(a) $\langle V^a_{||}\rangle$ (red) and $\langle V^b_{||}\rangle$ (blue)
  vs $F_{D}$ for the system from Fig.~\ref{fig:1} with
  $\alpha^a_m=0$, $\alpha^a_d=1.0$, and
  $\alpha^b_m=0.3$.
  (b) The corresponding $\langle V^a_{\perp}\rangle$  (red) and
  $\langle V^b_{\perp}\rangle$ (blue) vs $F_{D}$.
  (c) The corresponding $P_6$
  vs $F_{D}$.
  MC is the moving crystal state.
  There is a transition at $F_D=2.5$ from the moving liquid (ML) phase to
  a perpendicular stripe (PS) or cluster phase.}
\label{fig:5}
\end{figure}

In Fig.~\ref{fig:5}(a,b) we
plot $\langle V^a_{||}\rangle$,
$\langle V^b_{||}\rangle$,
$\langle V^a_{\perp}\rangle$,
and $\langle V^b_{\perp}\rangle$ versus
$F_{D}$ for the system in 
Fig.~\ref{fig:1} with $\alpha_m^a=0$, $\alpha_d^a=1.0$, and
$\alpha^{b}_{m} = 0.3$, while
in Fig.~\ref{fig:5}(c) we show the corresponding
$P_{6}$ versus $F_{D}$ curve.
Here we consider values of $F_D$ that are much higher than those presented in
Fig.~\ref{fig:1} in order to access the
transition from the ML state to a phase separated (PS) cluster state. 
The ML phase ends at $F_D=2.5$, where we find
a jump of $\langle V^a_{\perp}\rangle$ to $\langle V^a_{\perp}\rangle=0$,
indicating that the motion of species $a$ has locked to the $x$ direction, parallel
to the applied driving force.
This jump coincides with a jump in
$\langle V^b_{\perp}\rangle$ to more negative values,
and both curves are much smoother above the jump, indicating
that fluctuations are reduced in the PS state compared to the ML flow.
The ML-PS transition
is also associated with
a jump up in $P_{6}$ from $P_6=0.55$ in the ML phase to $P_6=0.9$ in the PS phase
as the system becomes more ordered into
a moving perpendicular stripe.

\begin{figure}
\includegraphics[width=3.5in]{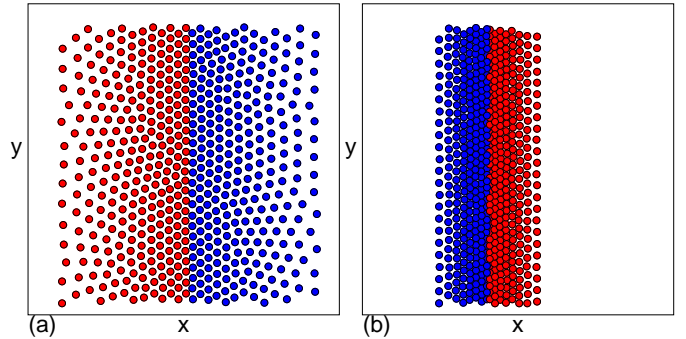}
\caption{
  Images of particle positions for species $a$ (red) and $b$ (blue).
  (a) At $F_{D} = 4.0$
  for the system in Fig.~\ref{fig:5} with $\alpha_d^a=1.0$, $\alpha_m^a=0$,
  and $\alpha^b_m=0.3$, 
  a perpendicular stripe (PS) state forms
  with a conformal crystal structure on each side.
  (b) The PS state for a system with $\alpha_d^a=1.0$, $\alpha_m^a=0$, and
  $\alpha^{b}_{m} = 0.75$ at
  $F_{D} = 6.0$, where the stripe is more compressed. }
\label{fig:6}
\end{figure}

In Fig.~\ref{fig:6}(a) we show the particle configurations from
the sample in Fig.~\ref{fig:5}(a) at $F_{D} = 4.0$
in the PS phase, where the particles
undergo species and density phase separation into
a partially clustered state consisting of a stripe aligned in the $y$ direction,
perpendicular to $F_D$.  The particle density is highest at the center of
the stripe.
Species $a$
moves only along the $x$-direction, causing it to
pile up behind species $b$,
which is moving in both the positive $x$ and negative $y$ directions.  As a result,
the entire pattern translates in the positive $x$ direction, and the two halves of the
pattern shear against each other along the $y$ direction.
In Fig.~\ref{fig:5}(a),
$\langle V^b_{||}\rangle$
is slightly higher than $\langle V^a_{||}\rangle$
in the ML phase, but in the PS phase
$\langle V^b_{||}\rangle$ and $\langle V^a_{||}\rangle$ become locked together.

There is considerable local sixfold ordering of the particles in the PS
state illustrated in,
Fig.~\ref{fig:6}(a),
but due to the density gradient the lattice is distorted into
conformal arch-like
patterns.
Conformal crystals arise in two-dimensional systems
of repulsive particles in the presence of some form of density gradient,
such as magnetic particles in a gravitational
field \cite{57},
vortices
in a Bean state \cite{58,59,60,61},
and colloidal particles under
a  gradient that is imposed by the system geometry \cite{62}.
The conformal crystals are the result of a competition between
the local sixfold ordering favored by
the repulsive particle-particle interactions
and the need to spatially vary the interparticle spacing in order
to accommodate the density gradient.
Most conformal crystals have been observed
under equilibrium conditions,
while the conformal crystal
structure illustrated in Fig.~\ref{fig:6}(a)
is a strictly
nonequilibrium state.
As $F_{D}$ increases,
the conformal stripe state becomes
more compressed along the $x$ direction.
Increasing $\alpha^b_m$ also compresses the PS stripe, as illustrated
in Fig.~\ref{fig:6}(b)
for
a system with $\alpha^a_m=0$, $\alpha_d^a=1.0$, and $\alpha^{b}_{m} = 0.7$
at $F_{D} = 5.0$.

\begin{figure}
\includegraphics[width=3.5in]{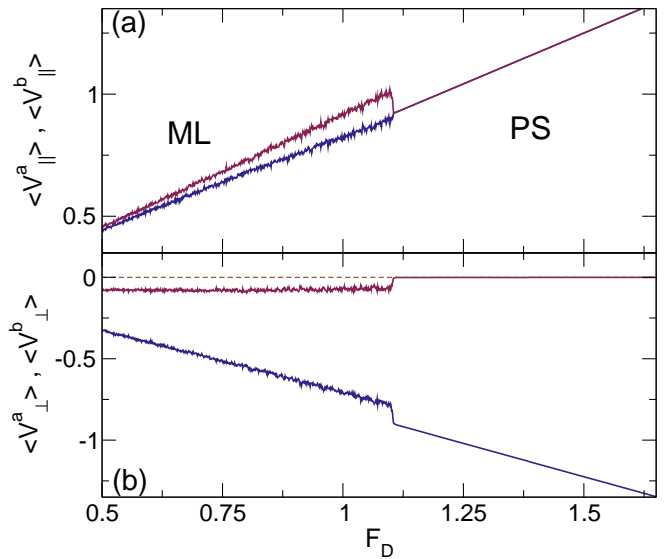}
\caption{(a) $\langle V^a_{||}\rangle$ (red) and $\langle V^b_{||}\rangle$ (blue)
  vs $F_{D}$ for a system with
  $\alpha^a_m=0$, $\alpha^a_d=1.0$, and $\alpha^b_m=0.7$ in the region on either side
  of the ML (moving lattice) to PS (perpendicular stripe) transition.
  (b) The corresponding $\langle V^{a}_{\perp}\rangle$ (red)
  and $\langle V^b_{\perp}\rangle$ (blue) vs $F_{D}$
  curves.  The dashed line indicates $\langle V_{\perp}\rangle=0$.
  The ML-PS transition is accompanied by a locking of the parallel velocities
  and a jump in the perpendicular velocities.}
\label{fig:7}
\end{figure}

In Fig.~\ref{fig:7}(a) we plot
$\langle V^{a}_{||}\rangle$ and $\langle V^{b}_{||}\rangle$
versus $F_{D}$ for the system in Fig.~\ref{fig:6}(b) with
$\alpha^b_m=0.7$ in the vicinity of the ML-PS transition,
while in 
Fig.~\ref{fig:7}(b) we show
the corresponding $\langle V^{a}_{\perp}\rangle$ and
$\langle V^{b}_{\perp}\rangle$ versus $F_{D}$
curves.
There is a clear parallel velocity locking,
with $\langle V^{a}_{||}\rangle/\langle V^{b}_{||}\rangle = 1.0$ in the PS state,
while $\langle V^{a}_{\perp}\rangle$ locks to zero velocity at the transition
at the same time as 
a jump in $\langle V^b_{\perp}\rangle$ to a more negative value appears.
The velocity fluctuations are reduced in the more ordered PS phase compared
to the disordered ML state.

\begin{figure}
\includegraphics[width=3.5in]{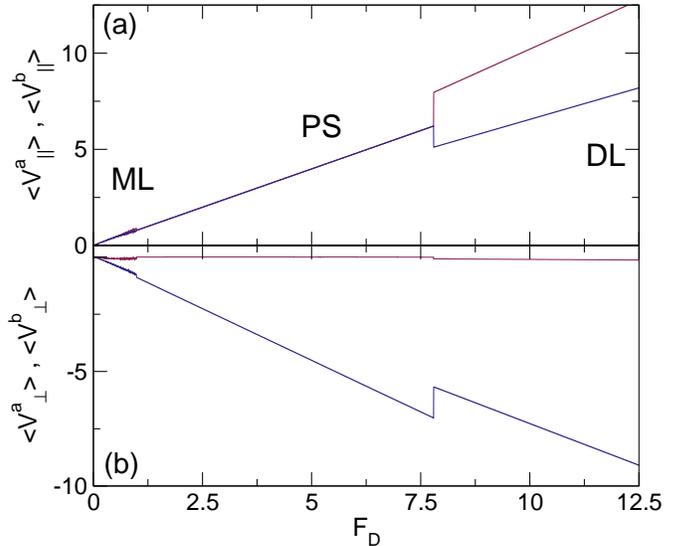}
\caption{ (a) $\langle V^a_{||}\rangle$ (red) and
  $\langle V^b_{||}\rangle$ (blue) vs $F_D$
  for a system with
  $\alpha^a_m=0$, $\alpha^a_d=1.0$, and $\alpha^{b}_{m} = 0.75$ showing
  that the ML (moving lattice) to PS (perpendicular stripe) transition
  is followed by a second transition to a diagonal laned (DL) state near
  $F_{D} = 7.75$ that is accompanied by large velocity jumps.
  (b) The corresponding $\langle V^a_{\perp}\rangle$ (red) and
  $\langle V^b_{\perp}\rangle$ (blue) vs $F_D$ curves.}
\label{fig:8}
\end{figure}

\begin{figure}
\includegraphics[width=3.5in]{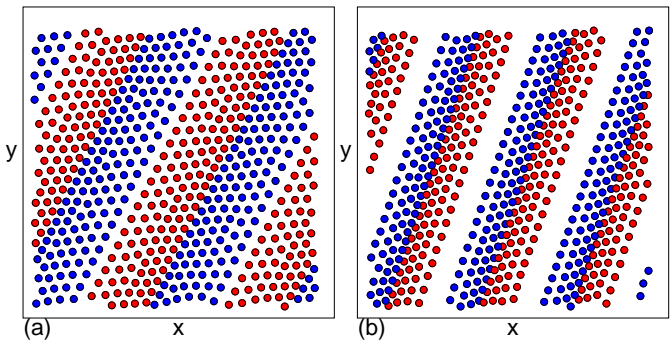}
\caption{
  Images of particle positions for species $a$ (red) and $b$ (blue) for
  the diagonal laned (DL) state.
  (a) The system in Fig.~\ref{fig:8}
  with $\alpha^a_d=1.0$, $\alpha^a_m=0$, and $\alpha^b_m=0.75$ at $F_D=8.0$.
  (b)
  A system with
  $\alpha^a_d=1.0$, $\alpha^a_m=0$, and
  $\alpha^{b}_{m} = 0.85$, showing the increased compression
of the diagonal stripes.}
\label{fig:9}
\end{figure}

When $F_D$ or
$\alpha^{b}_{m}$ is increased,
the stripes become more compressed and
the particle-particle interactions 
become strong enough to generate an instability that causes the
system to enter
a different phase consisting of multiple lanes with a more uniform
particle density.
In Fig.~\ref{fig:8} we
plot $\langle V^{a}_{||}\rangle$, $\langle V^b_{||}\rangle$,
$\langle V^a_{\perp}\rangle$, and $\langle V^b_{\perp}\rangle$
versus $F_D$ for a system with
$\alpha^b_{m} = 0.75$.
There is a large jump in the velocities near $F_{D} = 7.75$,
where the system undergoes
a transition from the
PS state to what we term a diagonal laned (DL) state.
As illustrated
in Fig.~\ref{fig:9}(a) for $F_{D} = 8.0$,
this state is composed of
multiple stripes of particles oriented at an angle to the driving direction.
Within each stripe, the density gradient is reduced compared to what is
observed in the PS state.
At the transition into the DL state,
$\langle V^a_{||}\rangle$ jumps up since species $b$ no longer blocks
the motion of species $a$ along the $x$ direction.
At the same time, there is a downward
jump in
$\langle V^{b}_{||}\rangle$
since species $b$ is no longer being pushed as hard in the $x$ direction
by species $a$.
Interestingly, 
a small downward jump in $\langle V^{b}_{\perp}\rangle$ occurs at the PS-DL
transition, 
indicative of negative differential conductivity.
This jump
is
produced by a Magnus force induced velocity
imparted by species $a$ to species $b$.
In the PS phase, species $a$ pushes against species $b$ along the $x$ direction,
and due to the Magnus term,
additional velocity
components arise for species $b$ 
in the negative $y$-direction.
In the DL state,
the tilt of the stripes
diminishes the magnitude of the $x$-direction push
from species $a$ on species $b$, which reduces the Magnus force
contribution to the $y$-direction velocity of species $b$.
At the same time, species $a$ now has a negative $y$ direction component of
force on species $b$ that generates a negative $x$ direction velocity component
of species $b$ through the Magnus term.
This produces the drop in $\langle V^b_{||}\rangle$ at the PS-DL transition. 
As $\alpha^b_m$ increases,
the PS-DL transition shifts
to lower values of $F_{D}$,
and when $\alpha^{b}_{m} \geq 0.85$,
the system passes directly from the ML to the DL phase without forming
a PS state.
In Fig.~\ref{fig:9}(b) we illustrate the DL state 
at $\alpha^b_m=0.85$ and $F_{D} = 8.0$, where the number of stripes
has increased.
Within the DL phase,
the stripes
become
more compressed
as $F_{D}$ increases until
a transition occurs to a new diagonal laned state, DL$_2$,
containing a larger number of stripes with fewer rows of particles in each stripe.

\begin{figure}
\includegraphics[width=3.5in]{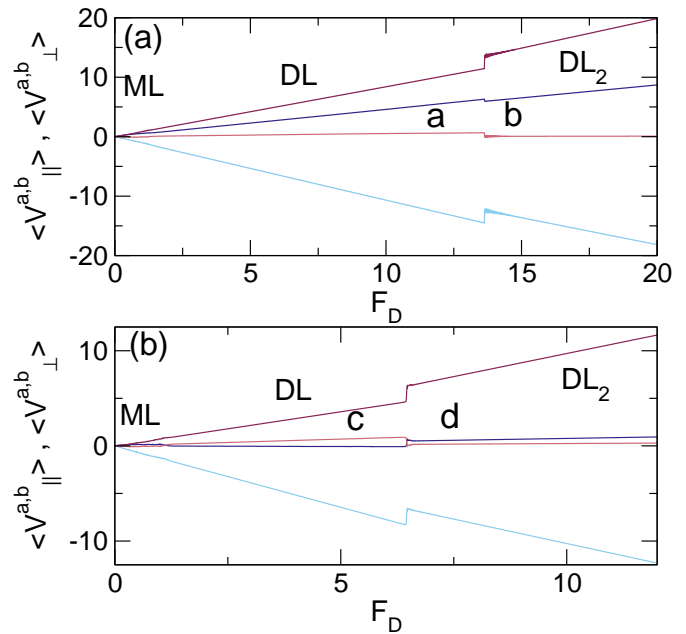}
\caption{(a) $\langle V^a_{||}\rangle$ (dark red),
  $\langle V^b_{||}\rangle$ (dark blue),
  $\langle V^a_{\perp}\rangle$ (pink),
  and $\langle V^b_{\perp}\rangle$ (light blue)
  vs $F_D$
  for a system with
  $\alpha^a_m=0$, $\alpha^a_d=1.0$, and
  $\alpha^{b}_{m} = 0.9$.
  The system passes directly from the ML (moving lattice) state to the DL (diagonal laned)
  state, followed by a transition to the DL$_2$ (second diagonal laned) state
  near $F_{D} = 13.5$.
  (b)
  The same for a system with $\alpha^a_m=0$, $\alpha^a_d=1.0$, and
  $\alpha^{b}_{m} = 0.998$.
The letters a, b, c, and d indicate the values of $F_D$ at which
the images in Fig.~\ref{fig:11} were obtained.
}  
\label{fig:10}
\end{figure}

\begin{figure}
\includegraphics[width=3.5in]{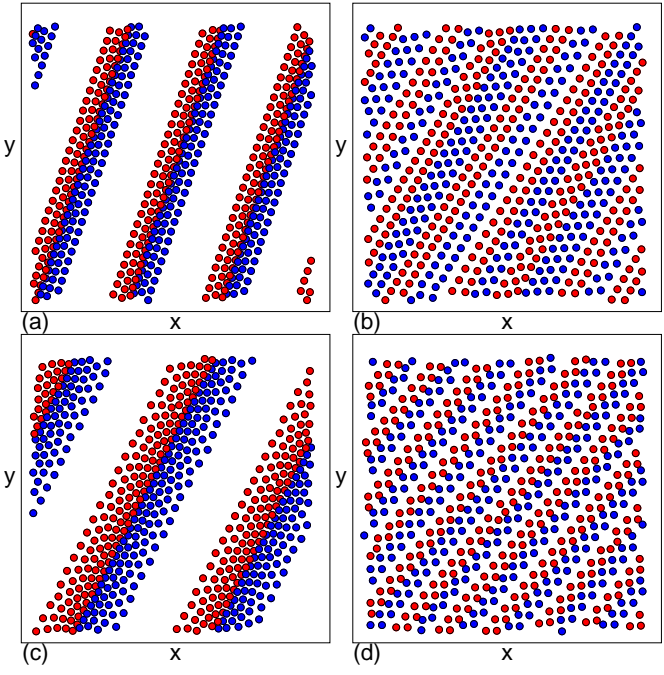}
\caption{
  Images of particle positions for species $a$ (red) and $b$ (blue).
  (a) The DL state for the system in Fig.~\ref{fig:10}(a)
  with $\alpha_d^a=1.0$, $\alpha_m^a=0$, and $\alpha_m^b=0.9$
  at $F_{D} = 10.35$.
  (b) The DL$_2$ state from Fig.~\ref{fig:10}(a) at $F_{D} = 14.75$. 
  (c) The DL state from the system in Fig.~\ref{fig:10}(b)
  with $\alpha_d^a=1.0$, $\alpha_m^a=0$, and $\alpha_m^b=0.998$
  at $F_{D} = 5.0$.
  (c) The DL$_2$ state from the system in Fig.~\ref{fig:10}(b) at $F_{D} = 7.5$.}   
\label{fig:11}
\end{figure}

In Fig.~\ref{fig:10}(a) we plot
$\langle V^a_{||}\rangle$,
$\langle V^b_{||}\rangle$,
$\langle V^a_{\perp}\rangle$,
and $\langle V^b_{\perp}\rangle$ versus $F_D$
for 
a system with $\alpha^b_{m} = 0.9$,
which transitions directly from the ML to the DL phase.
A second transition appears
near $F_{D} = 13.5$, as
indicated by the jumps in the velocity curves,
which corresponds to a rearrangement into the more uniform
laned state DL$_2$.
In Fig.~\ref{fig:11}(a) we
illustrate the DL state for the system in
Fig.~\ref{fig:10}(a) at $F_{D} = 10.3$, where the particles 
form a series of diagonal stripes,
each of which is composed of three rows of each species on
either side.
At $F_{D} = 14.75$ in the DL$_2$ phase,
as shown in Fig.~\ref{fig:11}(b),
there are still two to three rows
of particles on each side of each stripe, but the
stripes are much more spread out so that the particle density is
considerably more uniform.
In Fig.~\ref{fig:10}(b),
we plot 
$\langle V^a_{||}\rangle$,
$\langle V^b_{||}\rangle$,
$\langle V^a_{\perp}\rangle$,
and $\langle V^b_{\perp}\rangle$ versus $F_D$
for a system with $\alpha^b_m=0.998$ where $\alpha^b_{m}/\alpha^{b}_{d} = 15.79$,
in which the same phases appear but the DL-DL$_2$ transition is shifted
to a lower drive of $F_{D} = 6.5$.  
Figure~\ref{fig:11}(c) shows an image
of the DL phase for the system in
Fig.~\ref{fig:10}(b) at $F_{D} = 5.0$, where dense tilted stripes appear,
each of which contains
four to five rows of particles on each side.
In Fig.~\ref{fig:11}(d), the DL$_2$ phase for the same system
at $F_{D} = 7.5$
has a larger number of lower density stripes containing two rows of particles on
each side, giving a more uniform particle density.
The exact particle configurations
in the DL states are not unique,
but generally we find a small number of diagonal stripes with three or more rows
of particles on each side of each stripe in the DL phase, while
the DL$_2$ phase
is more uniform and each of the more numerous stripes has two to three rows of
particles on each side.
As $F_{D}$ is further increased,
the stripes in the DL$_2$ phase become more compressed,
and an additional transition can occur
to a DL$_3$ state in which each stripe has only a single row of particles
on each side.

\begin{figure}
\includegraphics[width=3.5in]{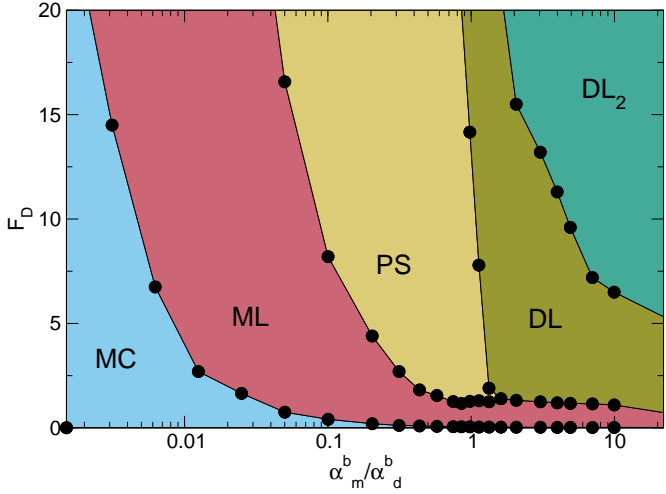}
\caption{Dynamic phase diagram as a function of $F_{D}$
  vs $\alpha^{b}_{m}/\alpha^b_d$ for the system in Fig.~\ref{fig:10}
with $\alpha^{a}_{d} = 1.0$ 
and $\alpha^{a}_{m} = 0$.
MC: moving crystal; ML: moving liquid; PS: perpendicular stripe;
DL: diagonal laned phase; DL$_2$: second diagonal laned phase.
} 
\label{fig:12}
\end{figure}

From the features in the transport curves
and changes in the particle configurations, we
construct the dynamical phase diagram 
shown 
in Fig.~\ref{fig:12}
as a function of $F_{D}$ versus
$\alpha^{b}_{m}/\alpha^b_d$ for samples with fixed
$\alpha^{a}_{d} = \alpha^{b}_{d} = 1.0$
and $\alpha^{a}_{m} = 0$.
The extent of the MC phase
increases as  $\alpha^{b}_{m}/\alpha^b_d$ goes to zero,
and similarly the ML-PS transition shifts to higher $F_D$ as
$\alpha^{b}_{m}/\alpha^b_d$ decreases.
The PS phase
only occurs when $\alpha^{b}_{m}/\alpha^b_d < 0.85$,
while for higher values of $\alpha^{b}_{m}/\alpha^b_d$
the system transitions directly from the ML phase to the DL state.
The DL$_2$ state appears when
$\alpha^{b}_{m}/\alpha^b_d \geq 0.9$. 

\section{Varied Damping Ratios}

\begin{figure}
\includegraphics[width=3.5in]{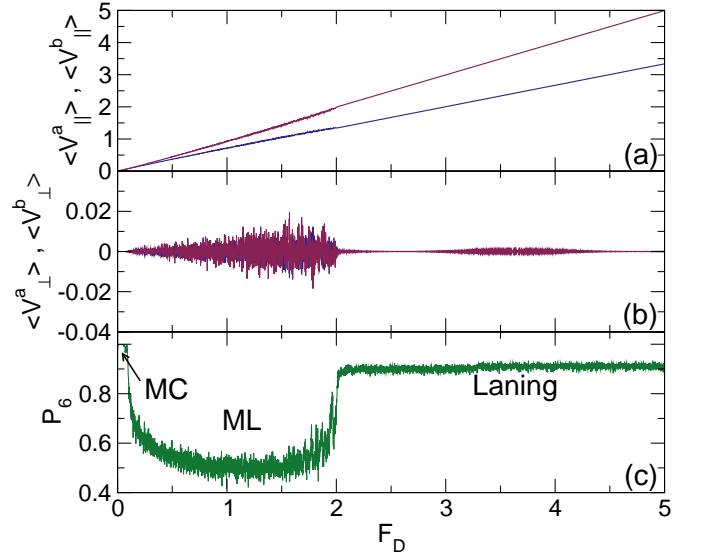}
\caption{(a) $\langle V^a_{||}\rangle$ (red) and $\langle V^b_{||}\rangle$ (blue)
  vs $F_{D}$ for a system with $\alpha^{a}_m=\alpha^b_{m} = 0$,
  $\alpha^{a}_{d} = 1.0$, and $\alpha^{b}_{d} = 2.0$.
  (b) The corresponding $\langle V^a_{\perp}\rangle$ (red)
  and $\langle V^b_{\perp}\rangle$ (blue) vs $F_{D}$, which
  both fluctuate around zero but
 show
  a clear change in the magnitude of the
  fluctuations
  near $F_{D} = 2.1$.
  (c) The corresponding $P_{6}$ vs $F_{D}$ 
showing the moving crystal (MC), moving liquid (ML), and laning states.} 
\label{fig:13}
\end{figure}

\begin{figure}
\includegraphics[width=3.5in]{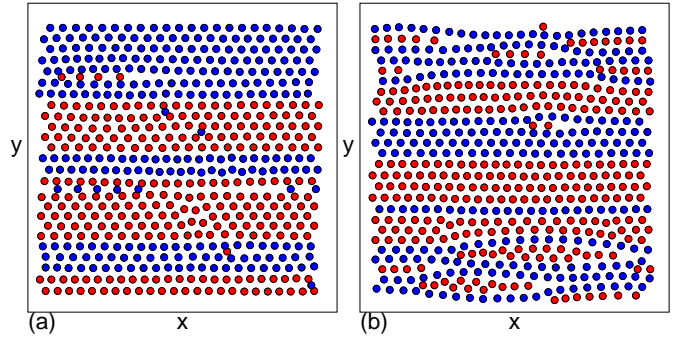}
\caption{
  Images of particle positions for species $a$ (red) and $b$ (blue).
  (a)
  The system in Fig.~\ref{fig:13}
  with $\alpha^a_m=\alpha^b_m=0$, $\alpha_d^a=1.0$, and $\alpha_d^b=2.0$
  in the laned state at $F_{D} = 2.5$.
  (b)
  The laned state for a system with
  $\alpha^a_m=\alpha_m^b=0$, $\alpha_d^a=1.0$, and
  $\alpha^{b}_{d} = 10$ at $F_D=2.5$.}  
\label{fig:14}
\end{figure}

\begin{figure}
\includegraphics[width=3.5in]{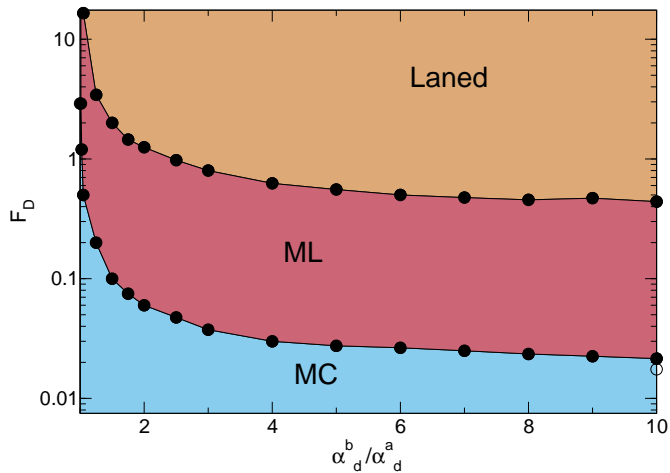}
\caption{Dynamic phase diagram as a function of $F_D$ vs $\alpha^a_d/\alpha^b_d$ for
  the system in Figs.~\ref{fig:13} and \ref{fig:14} with $\alpha^a_m=\alpha^b_m=0$
  and $\alpha_d^a=1.0$, showing the
  moving crystal (MC), moving liquid (ML), and laned states.
}
\label{fig:15}
\end{figure}

We next consider systems in which
$\alpha^{a}_{m}$ and $\alpha^{b}_{m}$ are both set equal to 
zero as the damping constant ratio is varied.
We fix $\alpha^a_{d} = 1.0$ and
vary $\alpha^b_{d}$.
This
is very similar to
systems of oppositely driven particles \cite{27},
but here
both species are moving in the same direction.
In Fig.~\ref{fig:13}(a,b) we plot
$\langle V^a_{||}\rangle$,
$\langle V^b_{||}\rangle$,
$\langle V^a_{\perp}\rangle$, and
$\langle V^b_{\perp}\rangle$,
for a system with
$\alpha^{a}_{d} = 2.0$.
Since the Magnus force is zero,
$\langle V^a_{\perp}\rangle$ and $\langle V^b_{\perp}\rangle$ both
fluctuate around zero.
At low drives,
$\langle V^a_{||}\rangle$ and
$\langle V^b_{||}\rangle$ are locked together
and the system 
forms an elastic triangular
solid in the moving crystal state.
As the drive increases, there is a transition to a moving liquid phase
in which the species $a$ particles, which have
a smaller damping coefficient,
move faster in the direction of drive 
than the species $b$ particles, which have higher damping.
At $F_{D} \approx 2.1$,
there is an abrupt decrease in the magnitude of the fluctuations
in both $\langle V^a_{\perp}\rangle$ and $\langle V^b_{\perp}\rangle$
as well as a cusp in
$\langle V^a_{||}\rangle$ and $\langle V^b_{||}\rangle$,
above which the velocities parallel to the drive increase linearly with drive.
In Fig.~\ref{fig:13}(c) we
plot $P_{6}$ for all the particles versus $F_{D}$,
where we find
that
$P_{6} = 1.0$
in the ordered MC phase.
At the transition to the ML phase, $P_6$ drops,
but for drives above
$F_D=2.1$,
there is a recovery of order to $P_{6}\approx 0.9$ 
as the system
enters a uniform laned state of the type
illustrated in Fig.~\ref{fig:14}(a) for $F_{D} = 2.5$.
This laned state is the same as that found
for oppositely driven Yukawa particles \cite{19},
and it is relatively ordered due to the
large patches of triangular lattice
that appear.
Sliding of adjacent lanes past each other is made possible by
the presence of aligned
5-7 dislocation pairs that can glide parallel to the driving direction,
which give the state a smectic character.
The zero Magnus term laned state
is distinct 
from the DL and DL$_2$ states
that appear for a finite Magnus term
in that the lanes are aligned with the driving
direction
and the particle density is uniform for all values of $F_{D}$ and
$\alpha^b_{d}$, unlike the
DL and DL$_2$ states,
which can show strong density modulations.
In Fig.~\ref{fig:14}(b) we illustrate that the laned state for $\alpha^b_{d} = 10$
has features similar to the laned states that form at lower $\alpha^{b}_{d}$.
In Fig.~\ref{fig:15} we construct a dynamic phase diagram as a function of
$F_{D}$ versus $\alpha^b_{d}/\alpha^{a}_{d}$,
and highlight the MC, ML, and 
laned states.
As $\alpha^{b}_{d}/\alpha^{a}_{d} \rightarrow 1$,
there is a divergence in the value of
$F_{D}$ at which the MC-ML transition occurs, along with a
similar divergence of the transition from the ML to the laned state.

\section{Varied Magnus Force}

\begin{figure}
\includegraphics[width=3.5in]{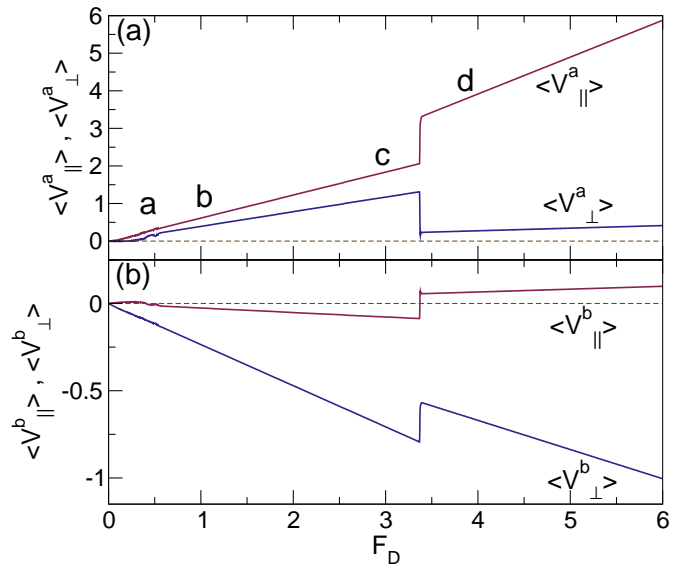}
\caption{(a) $\langle V^a_{||}\rangle$ (red) and $\langle V^a_{\perp}\rangle$ (blue)
  vs $F_{D}$ for a system with $\alpha_d^a=\alpha_d^b=1.0$,
  $\alpha_m^a=0$, and $\alpha_m^b=6.0$.
  (b) The corresponding $\langle V^{b}_{||}\rangle$ (red)
  and $\langle V^{b}_{\perp}\rangle$ (blue) vs $F_{D}$.
  The dashed lines indicate zero velocity,
  showing that there is a regime in which
  $\langle V^{b}_{||}\rangle$ becomes more negative as $F_D$ increases,
  meaning  that species $b$ is exhibiting 
  absolute negative mobility.}   
\label{fig:16}
\end{figure}

\begin{figure}
\includegraphics[width=3.5in]{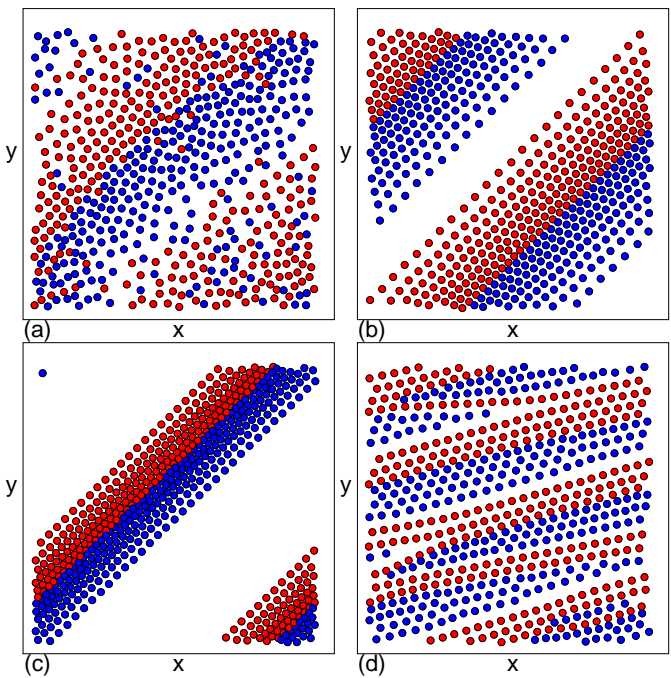}
\caption{Images of particle positions for species $a$ (red) and $b$ (blue)
  for the system in Fig.~\ref{fig:16} with
  $\alpha_d^a=\alpha_d^b=1.0$, $\alpha^a_m=0$, and $\alpha^{b}_{m} = 6.0$.
  (a) The moving liquid state at $F_D=0.5$, just before the transition into the
  diagonal laned (DL) state.
  (b) The DL state at $F_{D} = 1.25$.
  (c) The DL state at $F_{D} =3.0$ where the stripes are more compressed.
  (d) The DL$_2$ state at $F_{D} = 3.75$.} 
\label{fig:17}
\end{figure}

We next
relax the constraint $\alpha_{m}^b+  \alpha_{d}^b = 1.0$
and hold $\alpha^a_d=\alpha^b_d=1.0$ while fixing
$\alpha^{a}_{m} = 0.0$ 
and increasing $\alpha^{b}_{m}$ from
$\alpha^b_m=0$ to $\alpha^b_m=40$.
This protocol produces
more pronounced differences in the 
longitudinal and transverse velocities of the two species.   
In Fig.~\ref{fig:16}(a) we plot
$\langle V^a_{||}\rangle$ and $\langle V^a_{\perp}\rangle$ versus $F_D$
for a system with $\alpha^b_m=6.0$,
and in Fig.~\ref{fig:16}(b) we show the corresponding
$\langle V^b_{||}\rangle$ and $\langle V^b_{\perp}\rangle$ versus $F_D$.
When
$F_{D} < 0.6$,
we observe a disordered or partially disordered state,
as illustrated in Fig.~\ref{fig:17}(a) at
$F_{D} = 0.5$, where
a diagonal stripe is beginning to form.       

For $F_{D} > 0.6$ the system enters a DL state
as shown in Fig.~\ref{fig:17}(b) for
$F_{D} = 1.25$,
where the lanes are aligned at an angle of roughly
$\theta_l=60^\circ$
to the driving direction.
Within the DL state,
$\langle V^{a}_{||}\rangle$ and $\langle V^{a}_{\perp}\rangle$
are both positive,
indicating that species $a$ is sliding along the positive $x$ and positive $y$ directions
due to the orientation of the lane.
In contrast, $\langle V^{b}_{||}\rangle$ and $\langle V^{b}_{\perp}\rangle$
are both negative,
indicating that 
species $b$ is
moving
{\it opposite} to the direction of the applied drive, a phenomenon that is known as
absolute
negative mobility \cite{63,64,65,66}.
The strong Magnus force of the species $b$ particles  rotates the
$F_D$ component of the velocity
mostly into the negative $y$ direction,
leaving a residual positive $x$ direction velocity.
The overdamped species $a$ particles move parallel to the driving direction and
pile up behind the species $b$ particles, exerting a force of
magnitude $F_{ss}$ on them in both the
positive $x$ and negative $y$ directions with components
$F_{ss}\cos{\theta_l}$ and $F_{ss}\sin{\theta_l}$, respectively.
Under the Magnus force rotation,
the $F_{ss}\cos{\theta_l}$ portion of the particle-particle interaction
produces a species $b$ velocity contribution that is mostly in the negative
$y$ direction, while the $F_{ss}\sin{\theta_l}$ portion of the interaction
gives a velocity contribution that is mostly in the {\it negative} $x$ direction.
Since the positive $x$ direction velocities from $F_D$ and $F_{ss}\cos{\theta_l}$
are small, the net value of $\langle V^b_{||}\rangle$ is negative, resulting in
the absolute negative mobility that we observe.
As $F_{D}$ increases, the drift velocity of the entire diagonal lane in the positive
$x$ direction increases due to the increase in $\langle V^a_{||}\rangle$, but at the
same time the lane becomes more compressed, decreasing the distance between
the species $a$ and species $b$ particles at the center of the lane, and increasing
the particle-particle interaction force that is responsible for generating the negative
value of $\langle V^b_{||}\rangle$.  As a result, $\langle V^b_{||}\rangle$ becomes
more negative with increasing $F_D$.

We note that in previous studies of overdamped systems in which
negative mobility is observed,
the particles are
coupled to some type of asymmetric substrate
\cite{63,64,65,66}.
Magnus force-induced negative mobility
was predicted to occur for monodisperse skyrmions,
but only when the skyrmions are
coupled to a substrate \cite{67}.
The
bidisperse system we consider here
is unique in that a negative mobility can appear in the absence of a substrate.  

The compression of the diagonal lane becomes more intense with
increasing $F_D$, as illustrated in Fig.~\ref{fig:17}(c) at $F_D=3.0$
for the $\alpha_m^b=6.0$ system,
until a transition occurs at sufficiently high drive
to the more uniform DL$_2$ state
shown in Fig.~\ref{fig:17}(d) at $F_{D} = 3.75$.  
The DL-DL$_2$ transition is accompanied by
a jump up in $\langle V^a_{||}\rangle$ and
a drop in $\langle V^a_{\perp}\rangle$
since the lanes are now aligned closer to the driving direction at an
angle of $\theta_l=20^{\circ}$.
This change in lane orientation also causes
$\langle V^b_{||}\rangle$ to abruptly jump from a negative value to a positive
value at the DL-DL$_2$ transition.

\begin{figure}
\includegraphics[width=3.5in]{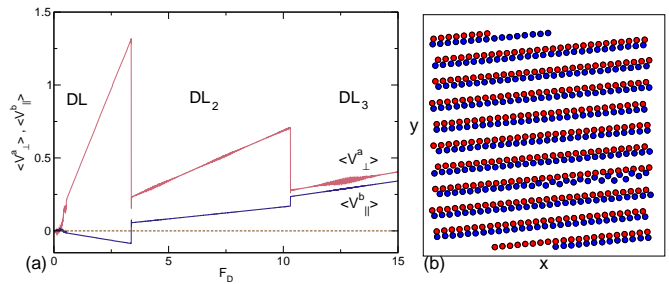}
\caption{$\langle V^{a}_{\perp}\rangle$ (pink)
  and $\langle V^{b}_{||}\rangle$ (blue)
  vs $F_{D}$ for the system in Fig.~\ref{fig:17}
  with $\alpha_d^a=\alpha_d^b=1.0$, $\alpha_m^a=0$, and $\alpha_m^b=6.0$
  showing jumps at the PS-DL,
  DL-DL$_2$, and DL$_2$-DL$_3$ transitions.
  (b) Images of particle positions for species $a$ (red)
  and $b$ (blue) in the DL$_3$ state at $F_{D} = 10.5$.}   
\label{fig:18}
\end{figure}

In Fig.~\ref{fig:18}(a) we plot
$\langle V^b_{||}\rangle$ and $\langle V^a_{\perp}\rangle$
versus $F_D$ for a sample with
$\alpha^{b}_{m} = 6.0$.
Here,
the negative value of $\langle V^b_{||}\rangle$
is more clearly visible.
At higher drives 
we find an
additional jump in both
$\langle V^b_{||}\rangle$
and $\langle V^a_{\perp}\rangle$
due to rearrangements
that transform the
DL$_2$ structure into what we call the DL$_3$ state,
illustrated
Fig.~\ref{fig:18}(b) for $F_D=10.5$.
The DL$_3$ lanes are very thin and are nearly aligned with the $x$ direction.

\begin{figure}
\includegraphics[width=3.5in]{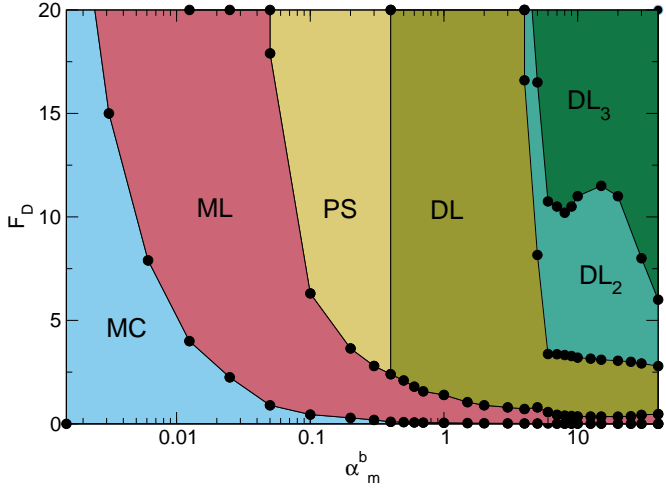}
\caption{Dynamic phase diagram as a function of
  $F_{D}$ vs $\alpha^{b}_{m}$ for a system with $\alpha^{a}_{m} = 0$
  and $\alpha^{a}_{d} = \alpha^{b}_{d} = 1.0$
  illustrating the moving crystal (MC), moving liquid (ML), perpendicular stripe (PS),
  and diagonal laned (DL, DL$_2$, and DL$_3$) phases.
A regime of negative mobility for species $b$ appears in phase DL.
}
\label{fig:19}
\end{figure}

In Fig.~\ref{fig:19} we construct a
dynamic phase diagram as a function of
$F_{D}$ versus $\alpha^b_{m}$ for the system in Figs.~\ref{fig:17} and \ref{fig:18}.
We find that the width of the
moving crystal
phase
diverges as $\alpha^b_{m} \rightarrow 0$,
while the PS phase
only occurs when $\alpha^{b}_{m} < 0.5$.
These results show that the phases we observe
are generic in systems where the damping coefficient is fixed but the
Magnus term varies.

\section{Magnus Terms of Opposite Sign}

\begin{figure}
\includegraphics[width=3.5in]{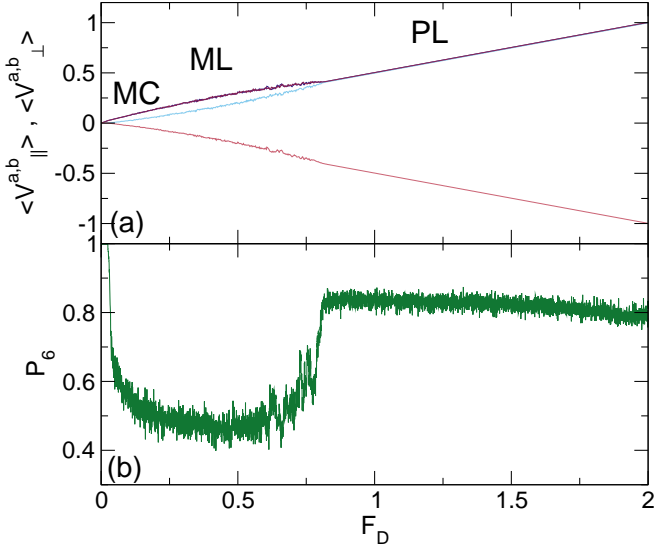}
\caption{(a) $\langle V^a_{||}\rangle$ (dark red),
  $\langle V^b_{||}\rangle$ (dark blue),
  $\langle V^a_{\perp}\rangle$ (pink), and
  $\langle V^b_{\perp}\rangle$ (light blue)
  vs $F_{D}$
  in a system with
  $\alpha^{a}_{m} = 1.0$, $\alpha^{b}_{m} = -1.0$,
  and $\alpha^{a}_{d} = \alpha^{b}_{d} = 1.0$.
  (b) The corresponding 
  $P_{6}$ vs $F_{D}$.
  At low drives we find a moving crystal (MC) state,
  followed by moving liquid (ML) and perpendicular laned (PL) phases.}
\label{fig:20}
\end{figure}

\begin{figure}
\includegraphics[width=3.5in]{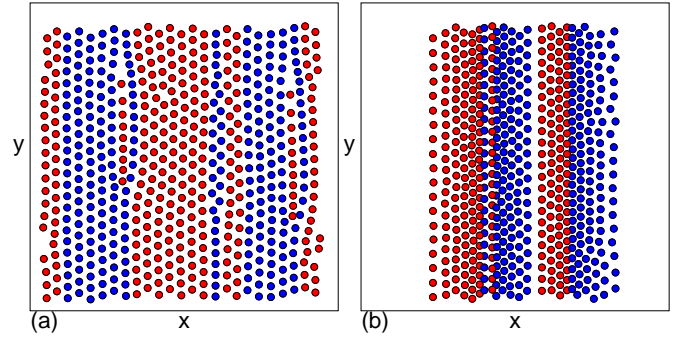}
\caption{Images of particle positions for species $a$ (red) and $b$ (blue).
  (a)
  The system in Fig.~\ref{fig:20} with $\alpha_m^a=1.0$,
  $\alpha_m^b=-1.0$, and $\alpha_d^a=\alpha_d^b=1.0$
  at $F_{D} = 1.5$
  showing the formation of a uniform
  perpendicular laned (PL) state
  oriented transverse to the
  driving direction.
  (b)
  A system with $\alpha^{a}_{m} = 1.0$, $\alpha^{b}_{m}=-1.0$, $\alpha^{a}_{d} = 1.0$, and
  $\alpha^b_{d} = 3.0$,
  in which the PL state shows compression or clustering.}     
\label{fig:21}
\end{figure}

Up to now we
have focused on
systems in which
the Magnus term is zero for one or both species, or where both species have a
finite Magnus force with the same sign but different magnitude.
Here we examine the case where
both species 
have a nonzero Magnus term of equal magnitude
that is opposite in sign.
In Fig.~\ref{fig:20}(a)
we plot
$\langle V^a_{||}\rangle$,
$\langle V^b_{||}\rangle$,
$\langle V^a_{\perp}\rangle$, and
$\langle V^b_{\perp}\rangle$
versus $F_{D}$
for a system with
$\alpha^{a}_{m} = 1.0$,
$\alpha^b_m=-1.0$, and $\alpha^{a}_{d} = \alpha^b_d=1.0$,
while in
Fig.~\ref{fig:20}(b)
we show the corresponding
$P_{6}$ versus $F_{D}$ curve.
Here $\langle V^a_{\perp}\rangle$  is negative while
$\langle V^a_{||}\rangle$, $\langle V^a_{\perp}\rangle$, and
$\langle V^b_{||}\rangle$ are all positive.
At low $F_D$ the system is in
an elastic MC state,
and it transitions with increasing drive
into a moving liquid in which
$P_{6} \approx 0.45$.
An ordering transition
occurs near $F_{D} = 0.75$,
as indicated by the increase in $P_{6}$ to
a value close to $P_6=0.865$,
and simultaneously
$\langle V^a_{||}\rangle$ and $\langle V^b_{||}\rangle$  lock with
$\langle V^b_{\perp}\rangle$
as the system enters
what we term
a perpendicular laned (PL) state,
illustrated
in Fig.~\ref{fig:21}(a) at $F_{D} = 1.5$.
The PL structure is very similar to
the lanes that form for
oppositely driven Yukawa particles,
except in this case the lanes are
oriented perpendicular to the drive. 
The PL state is distinguished from the PS state
by the fact that the density of the system remains uniform out to
arbitrarily high drives in the PL phase. 
In general, we find no significant density modulations
whenever $\alpha^{a}_{m}$ and $\alpha^b_m$ are of the same magnitude
but opposite in sign as long as $\alpha^a_d=\alpha^b_d$,
even when $\alpha^a_m$ or $F_D$ are very large.
By setting the damping terms of the two species to different values,
it is possible
to induce clustering in the PL state
with $\alpha^a_m=-\alpha^b_m$,
as illustrated in Fig.~\ref{fig:21}(b) for a system 
with $\alpha^{a}_{m} = 1.0$, $\alpha^b_{m} = -1.0$,
$\alpha^{a}_{d} = 1.0$, and $\alpha^{b}_{d} = 3.0$, 
where the perpendicular lanes are now compressed.
This compression arises because species $a$, which has a smaller damping term
than species $b$, moves faster
and collides with the band of slower moving particles,
while the opposite sign of the Magnus terms for the two species is
responsible for creating
the perpendicular banding.         

\begin{figure}
\includegraphics[width=3.5in]{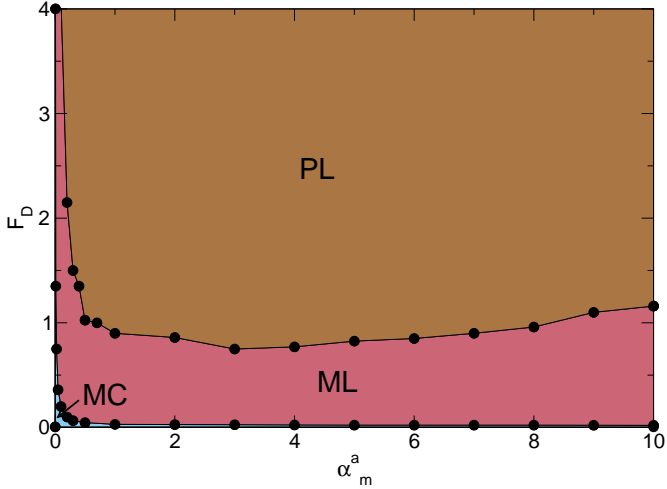}
\caption{Dynamic phase diagram as a function of $F_D$ vs $\alpha^a_m$
  for the system in Fig.~\ref{fig:20}
  with $\alpha_d^a=\alpha_d^b=1.0$ and
  $\alpha^b_m=-\alpha^a_m$
  showing the moving crystal (MC),
  moving liquid (ML), and perpendicular laned (PL) phase.}  
\label{fig:22}
\end{figure}

In Fig.~\ref{fig:22} we construct a
dynamic phase diagram as a function of $F_D$ versus $\alpha^a_m$
for the system in Fig.~\ref{fig:20} with $\alpha^b_m=-\alpha^a_m$,
highlighting the fact that the
widths of the MC and ML phases diverge
upon approaching the overdamped limit of $\alpha^a_m=0$.
The ML phase also increases in extent with increasing $\alpha^a_m$
when $\alpha^a_{m} > 5.0$.   

\section{Other Variables and Discussion}

\begin{figure}
\includegraphics[width=3.5in]{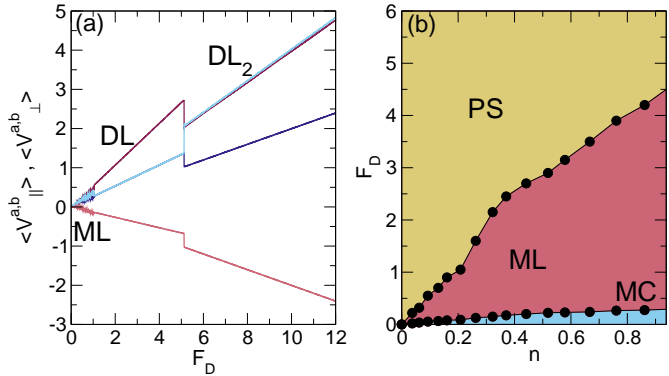}
  \caption{(a)
    $\langle V^a_{||}\rangle$ (dark red),
    $\langle V^b_{||}\rangle$ (dark blue),
    $\langle V^a_{\perp}\rangle$ (pink)
    and $\langle V^b_{\perp}\rangle$ (light blue)
    versus $F_D$
    for a system with
    $\alpha^{b}_{d} = 1.0$, $\alpha^{a}_{d} = 2.0$,
    $\alpha^{a}_{m} = 0.1$, and $\alpha^{b}_{m} = -1.0$, where
    a series of pronounced jumps indicate
    transitions into different
    DL and DL$_2$ phases. 
    (b) Dynamic phase diagram as a function of
    $F_{D}$ vs system density $n$ for
    samples with
    $\alpha^{a}_{d} = \alpha^{b}_{d} = 1.0$,
    $\alpha^{a}_{m} = 0$, and $\alpha^{b}_{m} = 0.3$,
    showing that the transitions between the moving crystal (MC),
    moving liquid (ML), and perpendicular stripe (PS) phases shift to
    higher values of $F_D$ with increasing density.
}
\label{fig:23}
\end{figure}

\begin{figure}
\includegraphics[width=3.5in]{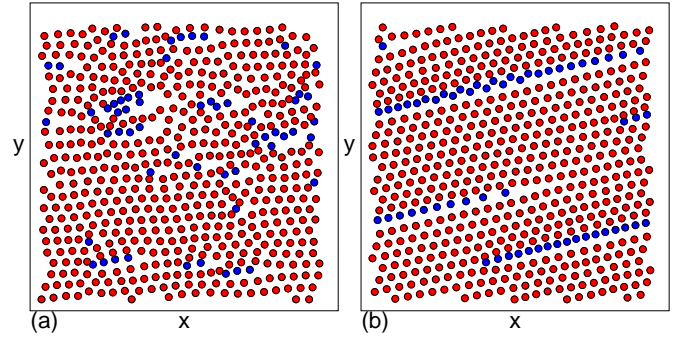}
\caption{Images of particle positions for species $a$ (red)
  and $b$ (blue) for a system with 
  $N_{b}/N_{a} = 0.9$,
  $\alpha^{a}_{d} = \alpha^{b}_{d}  = 1.0$,
  $\alpha^{a}_{m} = 0$, and $\alpha^{b}_{m} = 3.0$.
  (a) A moving liquid phase.
  (b) A diagonal stripe state,
  showing that phase separation persists even when the ratio of
  the number of species $a$ particles to species $b$ particles is varied.
}
\label{fig:24}
\end{figure}

We have considered several other variables
such as varied damping and varied Magnus ratios,
and in general find the same phases described above with some minor variations.
For example,
in Fig.~\ref{fig:23}(a) we plot
$\langle V^a_{||}\rangle$,
$\langle V^b_{||}\rangle$,
$\langle V^a_{\perp}\rangle$, and
$\langle V^b_{\perp}\rangle$ versus $F_D$
for a system with
$\alpha^{b}_{d} = 1.0$, $\alpha^{a}_{d} = 2.0$,
$\alpha^{a}_{m} = 0.1$, and $\alpha^{b}_{m} = -1.0$,
where we find
a series of pronounced velocity jumps at the transitions
into different DL and DL$_2$ phases.
The same generic phases persist when the system density is varied,
as shown in the dynamic phase diagram as a function of $F_D$ and $n$ in
Fig.~\ref{fig:23}(b) for a system with
$\alpha^{a}_{d} = \alpha^{b}_{d} = 1.0$,
$\alpha^{a}_{m} = 0$, and $\alpha^{b}_{m} = 0.3$.
As $n$ increases, the transitions between the phases
shift to higher values of $F_D$.
We observe a similar trend
for higher values of $\alpha^{b}_{m}$.
We have also examined systems in which $N_a \neq N_b$ and find similar
dynamic phases,
as illustrated in
Fig.~\ref{fig:24}(a,b) for a sample
with $N_{b}/N_{a} = 0.9$,
$\alpha^{a}_{d} = \alpha^{b}_{d}  = 1.0$,
$\alpha^{a}_{m} = 0$, and $\alpha^{b}_{m} = 3.0$,
where we show that in the ML phase,
some species segregation occurs,
while in the DL phase,  thinner stripes appear.
These results indicate that the general
features we observe are robust for a wide range of parameters. 

In this work we have only considered bidisperse particles;
however, it would also be interesting to study three or more
particle species
or even a continuum range of species with a Gaussian distribution of types.
In such assemblies, it is possible that
the system would generally
form  disordered or moving liquid
phases; however,
other new types of
pattern formation could appear.
We have only utilized particle-based simulations since
these 
allow us to 
simulate a large number of particles
over a wide range of parameters
for long times,
but it would also be interesting to perform continuum based simulations
of multiple species or sizes of skyrmions
to see whether similar effects arise when the internal degrees of freedom of
the skyrmions are included. 
In our system,
the Magnus term leads to the appearance of
a skyrmion Hall angle,
but there are recent studies
which show that some skyrmion systems
can exhibit motion that is more toroidal in addition
to other types of complex
dynamics \cite{N2,P}.
It would be interesting to study such systems in the presence of
multiple skyrmion species.
Further effects could arise if the applied driving were ac rather
than dc, since the multiple species could organize into different types of patterns
under cyclic driving.

\section{Summary}
We have examined a bidisperse system of particles
with uniform pairwise interactions under
dynamics that include both a damping term
and a Magnus term.
When both species have equal damping but only one species has a finite
Magnus term,
we find that the triangular lattice which forms under zero drive moves elastically
at low drives
with a Hall angle equal to the average Hall angle of the two species.
At a critical drive,
a Magnus induced disordering transition occurs
in which
each species moves with a different velocity in the direction perpendicular to the drive.
The critical drive
at which the disordering transition appears
diverges as the Magnus term goes to zero.
At higher drives,
there is a transition to a perpendicular stripe or cluster state
with both density and species phase separation.
The stripes become more compressed as the drive or difference in Magnus terms
increases,
and another transition occurs to
a density modulated diagonal laned state at even higher drives.
The
transitions
are associated with pronounced jumps and locking of
the transverse and longitudinal velocities of each species as well as
changes in the global particle structures.
We also find that
multiple transitions can occur within the diagonal laned phase,
each of which is accompanied by a reduction in both the number of particles in each row
and the angle between the stripe and the driving direction, 
giving rise to a rich variety 
of different types of patterns.
In some cases,
one of the species can
exhibit absolute negative mobility
in which the particles move in the direction opposite to that of the
applied drive due to the Magnus-induced rotation of the interaction forces between
the two particle species.
When both species have different damping terms but zero Magnus terms,
we find
dynamic phases that are very similar to those
observed for oppositely driven Yukawa particles,
which form uniform laned states.
For equal damping terms and Magnus terms that are equal in magnitude
but opposite in sign,
uniform density states appear containing
lanes that are perpendicular to the applied drive.
We show that the diagonal laned and density modulated states 
are robust for a wide range of parameters and densities,
and should be generic features of systems with dispersity in the Magnus force. 
We discuss the relation between 
our results
and studies of
skyrmion systems
with dispersion in the Magnus force,
where we predict that a disordering
transition should occur as a function of increasing drive, and that
a variety
of clustered and pattern forming states could be observed.
Similar effects may arise
in soft matter systems containing
Magnus terms, such as spinning magnetic particles in solution.         

\acknowledgments
This work was carried out under the auspices of the 
NNSA of the U.S. DoE at LANL under Contract No.
DE-AC52-06NA25396.

\end{document}